\documentclass[]{AO4ELT}  

\usepackage{microtype}
\usepackage{biblatex}
\usepackage{amsmath,amsfonts,amssymb}
\usepackage{graphicx}
\usepackage{pst-all} 
\usepackage[colorlinks=true, allcolors=blue]
{hyperref}
\usepackage{xcolor}
\makeatletter         
\def\@maketitle{
\includegraphics[width = 170mm]{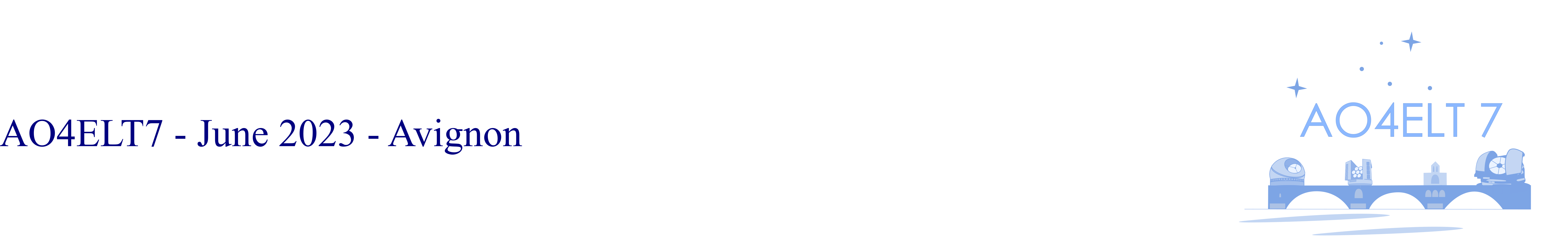}\\[8ex]
\begin{center}
{\Huge \bfseries \sffamily \@title }\\[4ex] 
{\Large  \@author}\\[4ex] 
\@date
\end{center}}

\title{Closed-Loop Until Further Notice: Comparing Predictive Control Methods in Closed-Loop}

\author[a]{J. Fowler}
\author[b]{M.A.M. van Kooten}
\author[a]{R. Jensen-Clem}
\affil[a]{Department of Astronomy \& Astrophysics, University of California, Santa Cruz, CA, USA}
\affil[b]{National Research Council Canada, Herzberg Astronomy and Astrophysics Research Center, Victoria, Canada}

\authorinfo{Further author information: (Send correspondence to J.F.)\\J.F.: E-mail: jumfowle@ucsc.edu}

\begin{document} 
\maketitle

\begin{abstract}
For future extremely large telescopes, error in extreme adaptive optics systems at small angular separations will be highly impacted by the lag time of the correction, which is typically on millisecond timescales; one solution is to apply a predictive correction to catch up with the system delay. Predictive control leads to significant RMS error reductions in simulation (on the order of 5-10x improvement in RMS error compared with a standard integral controller), but shows only modest improvement on-sky (less than 2x in RMS error). This performance limitation is likely impacted by elements of pseudo open loop (POL) reconstruction, which requires assumptions about the response of the deformable mirror and accuracy of the wavefront measurements that are difficult to verify in practice. In this work, we explore a closed-loop method for data-driven prediction using a reformulated empirical orthogonal functions (EOF). We examine the performance of the open and closed-loop methods in simulation on perfect systems and systems with an inaccurate understanding of the DM response. 

\end{abstract}

\keywords{predictive control, extreme adaptive optics (exAO), empirical orthogonal functions (EOF)}

\section{INTRODUCTION}
\label{sec:intro}  

Dessenne, 1998 \cite{Dessenne1998} was one of the first papers outlining predictive control for adaptive optics (AO) systems. This predictive method was demonstrated on-sky in 1999 \cite{Dessenne1999} at the 1.52 meter observatory in Haute Provence; they cite a relative Strehl increase over a classic
integrator of 30$\%$ (a max Strehl performance of $\sim 14\%$) at their visible central wavelength of 650 nm. Since then, a plethora of predictive control methods have come into the literature, including linear estimators \cite{Guyon2017, Guyon2018, jensen2019, vanKooten2022}, linear quadratic Gaussian controllers \cite{Paschall1993, Sivo2014, Poyneer2016, Petit2011}, model-based updates to a linear quadratic Gaussian \cite{Poyneer2007, Poyneer2023, Fowler2022}, subspace control methods \cite{Haffert2021}, and non-linear neural network solvers \cite{Liu2020, Wong2021, Landman2021, Swanson2021, Nousiainen2022, Hafeez2022}. Despite nearly 25 years of predictive methods, Empirical Orthogonal Functions (EOF)\cite{Guyon2017} is the only method to have been run on-sky as the controller for all spatial frequencies on an 8-10 meter class telescope; demonstrated on Subaru/SCExAO\cite{Guyon2018} and on Keck/NIRC2 \cite{vanKooten2022}. 

As opposed to more classic methods (e.g., an integral controller or a linear quadratic Gaussian controller) a classic EOF \cite{Guyon2017} learns a linear relationship in the evolution of the full scale of input turbulence, which means that EOF trains on (and predicts) the open loop wavefront. Because no single conjugate AO system runs in open loop, in practice these methods train on pseudo-open loop (POL) data, where deformable mirror (DM) commands are added to wavefront sensor (WFS) measurements to reconstruct the full state of turbulence. However, this leaves room for error in the reconstruction. Any spatially or temporally evolving mismatches in the calibration (e.g., non-linearities in the sensing and actuator pokes or DM to WFS misregistrations) that are not as impactful in an integral closed-loop controller could lead to incorrect performance prediction in open loop.  

However, upon closer inspection of Dessenne's adaptive predictor \cite{Dessenne1998} and Guyon's EOF \cite{Guyon2017}, it becomes clear that they are both methods that build predictive controllers from the same pieces of information. The major departure between the two methods is that Guyon's EOF runs in open loop, and reconstructs full turbulence states, and Dessenne's preditive controller runs in closed-loop, and encapsulates the evolution of both the DM commands and WFS measurements. With this work, we aim to revisit Dessenne's method, update it for comparison with Guyon's EOF, perform preliminary simulations of its feasibility within the context of a modern system, and ultimately begin to compare what is essentially an open and closed-loop implementation of empirical orthogonal functions. From hence forth, we will refer to both methods as EOF, and make the distinction between open and closed-loop implementations. 

In Section \ref{sec:POL_EOF}, we describe the pseudo-open loop EOF method, in Section \ref{sec:CL_EOF} we describe the closed-loop update for this work, and in Section \ref{sec:truth_conditions} we discuss training conditions for the closed-loop update. In Section \ref{sec:results} we present the results of preliminary performance simulations and the impact of model-mismatch errors on that performance. 

\section{EMPIRICAL ORTHOGONAL FUNCTIONS}

\begin{figure} [ht]
   \begin{center}
   \begin{tabular}{c} 
   \includegraphics[width=0.85\linewidth]{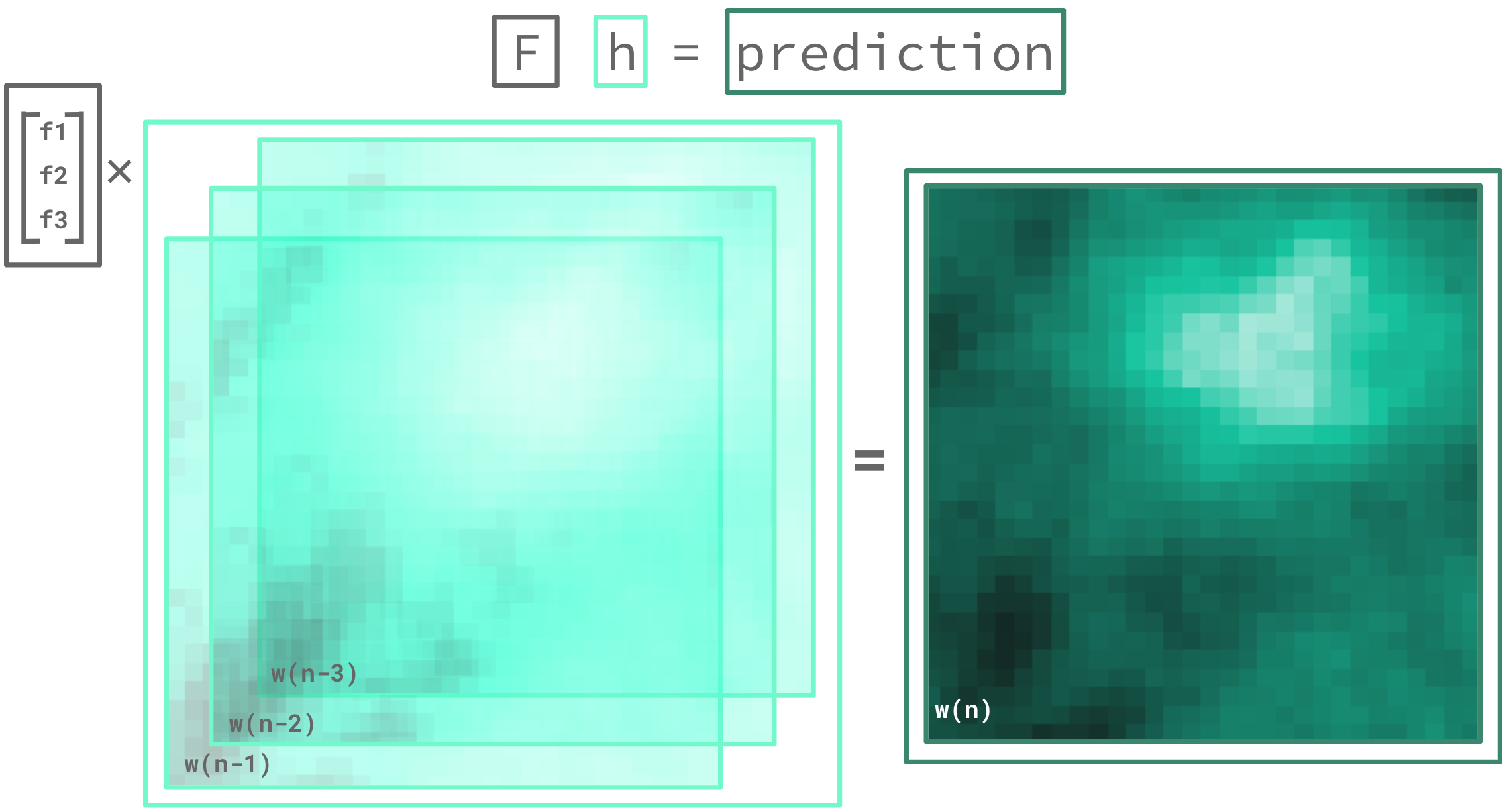}
   \end{tabular}
   \end{center}
   \caption[example] 
   { \label{fig:eof_diagram} 
The future state of the wavefront is predicted as a linear combination of previous states, $w$ at some time $n$. The collection of previous states is called the history vector, $\mathbf{h}$. Phase screens evolve in time (in sea foam) and a weight is applied to each one (the predictive filter $\mathbf{F} = [f_1, f_2, ...]$, where each $f_n$ contains m weights for m modes) to estimate the final wavefront prediction in teal. Note that mathematically a history vector is a single flat vector that contains the appended information of all 3 screens to predict a final vector that contains a flattened version of the predicted wavefront.}
   \end{figure}

Empirical orthogonal functions predicts a future state of a wavefront by building a predictive filter that linearly combines previous states of the wavefront. Figure \ref{fig:eof_diagram} shows a visual representation. In the following sections we describe the math that represents this process for open and closed-loop implementations. 

\subsection{Previous Open Loop Implementations}
{\label{sec:POL_EOF}}
Starting with Guyon, 2017 \cite{Guyon2017} and a follow up from Jensen-Clem in 2019 \cite{jensen2019}, we outline our implementation of EOF in (pseudo-)open loop.
Given some state of the wavefront (i.e., full scale of the uncorrected turbulence as considered in pseudo open loop), $w$, with m variables representing wavefront sensor measurements (which for a zonal approach corresponds to the number of  deformable mirror actuators) and n associated frames, we build a history vector for each subaperture:
 
\begin{equation}
\mathbf{h}_m(t) = 
\begin{bmatrix}
w(t) \\
w(t -dt) \\
w(t-(n-1)dt) 
\end{bmatrix}
\end{equation}

We build a predictive filter for each mode \textbf{F}$_m$ that will predict the full phase at a given point with: 

\begin{equation}
    \mathbf{F}_m \mathbf{h}_m(t) = w(t+dt)
\end{equation}

Both the filter and history vector can be written to include information across multiple modes, i.e., $\mathbf{h} = [w_0(t), w_1(t) ... w_0(t-dt), w_1(t-dt) ...]$, but for ease of comparison with the closed-loop implementation we leave these as single wavefront sensor measurement/DM actuator filters and predictions.

To find the predictive filter $\mathbf{F}$ we minimize an error term that consists of the difference between the output predicted wavefront (in DM space) and the true phase at that time. We collect training data $\mathbf{D}$, which contains history vectors and $\mathbf{P}$ which holds the true state (or ``future") for each history vector. (I.e., mapping $t$ to its future state at time $t+dt$.) 

\begin{equation}
\textrm{min}_\mathbf{F}||\mathbf{D}^T\mathbf{F}^T - \mathbf{P}^T||^2 
\end{equation}

Solving equation (3) requires a pseudo-inverse; Guyon, 2017 \cite{Guyon2017} solved this problem with an SVD (singular value decomposition) inversion, but we use a least-squares inversion \cite{jensen2019}, with regularization constant $\alpha$ ($\alpha$ may be set to $1$ for simulations, but is found empirically on-sky \cite{vanKooten2022}.)

\begin{eqnarray}
\mathbf{F} &=& ((\mathbf{D}^T)^\dagger \mathbf{P}^T)^T \\
\mathbf{F} &=& \mathbf{P}\mathbf{D}^T(\mathbf{D}\mathbf{D}^T + \alpha \mathbf{I})^{-1}
\end{eqnarray}

Finally, our predictive filter $\mathbf{F}$ holds a coefficient for each previous state, as expressed by a pseudo open loop (POL) reconstruction of the telemetry projected into DM space.

\subsection{Our Development of a Closed-Loop Update}

{\label{sec:CL_EOF}}

Inspired by Dessenne, 1998 \cite{Dessenne1998} and Haffert, 2021 \cite{Haffert2021}, we explore a closed-loop reformulation of the classic empirical orthogonal functions \cite{Guyon2017}. The fullest realization of this method could: 
\begin{enumerate}
    \item Improve controller stability by running in closed-loop. 
    \item Avoid error introduced by non-linear wavefront sensing or DM model mismatch when wavefront sensor residuals are converted to psudeo open loop. 
    \item Track drifts that may impact the DM and wavefront sensor separately by allowing each to evolve with a different set of coefficients. 
    \item Do robust timekeeping with information into the system. 
\end{enumerate}

Following \cite{Dessenne1998}, an updated  history vector (we refer to this as $\boldsymbol{\phi}$ to distinguish from the open loop history vector, as in standard in control theory conventions), contains both wavefront sensor residuals $\epsilon(t)$ and DM commands $y(t)$. (While the original work opted to use KL-modes, we work in the DM zonal basis, i.e., one point of information per DM actuator/WFS subaperture.) We consider only a single mode at a time: 

\begin{equation}
\boldsymbol{\phi}(n) = 
\begin{bmatrix}
y(n-1)\\
y(n-2)\\
... \\
y(n-p) \\
\epsilon(n-2)\\
\epsilon(n-3)\\
... \\
\epsilon(n-p-1)
\end{bmatrix}
\end{equation}

In this way, the new history vector and the corresponding predictive filter have twice as many values per mode pair as previous open loop derivations. However, the output of the filter applied to the history vector $\boldsymbol{\theta}\cdot\boldsymbol{\phi}$ predicts the same piece of information as an open loop implementation: the full turbulence in DM space, which is the DM command needed at a given iteration, in our notation $y(n)$. 

Dessenne's original formulation suggests reconstructing open loop turbulence from the full transfer function of the AO control loop to train the predictive filter, and solving for a steady state recursive least squares solution \cite{Dessenne1998}. In this simulation, we instead apply the same minimization technique used in the open loop implementation. Future work, inspired by that of van Kooten, 2019 \cite{vanKooten2019} will explore a recursive least square implementation that updates the filter with each control iteration. We build a new minimization problem: 

\begin{equation}
\textrm{min}_{\boldsymbol{\theta}}||\mathbf{D}^T\boldsymbol{\theta}^T - \mathbf{P}^T||^2 
\end{equation}

where $\mathbf{D}$ contains collections of the history vector $\boldsymbol{\phi}$, and $\mathbf{P}$ contains the future state of the full turbulence. 

In practice, the distinction between the open and closed-loop implementation is twofold: (1) the ability of the correction to apply different coefficients for the DM commands and the wavefront sensor information (whereas POL is one command applied to a summed value) and (2) the ability to robustly encapsulate a time delay into the reconstruction. POL reconstruction adds wavefront sensor measurements to DM commands with a single static delay, while closed-loop methods can account for when each piece of information enters the system (with two-step delays or delays with non-uniform steps). See Appendix \ref{sec:time-steps} for additional information on how timekeeping impacts the construction of $\boldsymbol{\phi}$ and $\mathbf{P}$.

Assuming a classic two-step delay and $p$ associated frames of information in a history vector, the weighted estimate for a single coefficient for the closed-loop formulation ($y_{CL}$) vs. the original open loop formulation ($y_{OL}$) takes the form: 
\begin{eqnarray}
   y_{CL}(n) &=& b_1y(n-1) + ... + b_{p}y(n-p) + a_0\epsilon(n-2) + ... + a_{p-1}\epsilon(n-p-1) \\
   y_{OL}(n) &=& c_{0}[y(n-1) + \epsilon(n-1)] + ... + c_p[y(n-p-1) + \epsilon(n-p-1)]
\end{eqnarray}

\subsection{Selecting a Truth Condition for Training Data}

\begin{figure} [ht]
   \begin{center}
   \begin{tabular}{c} 
   \includegraphics[height=0.34\linewidth]{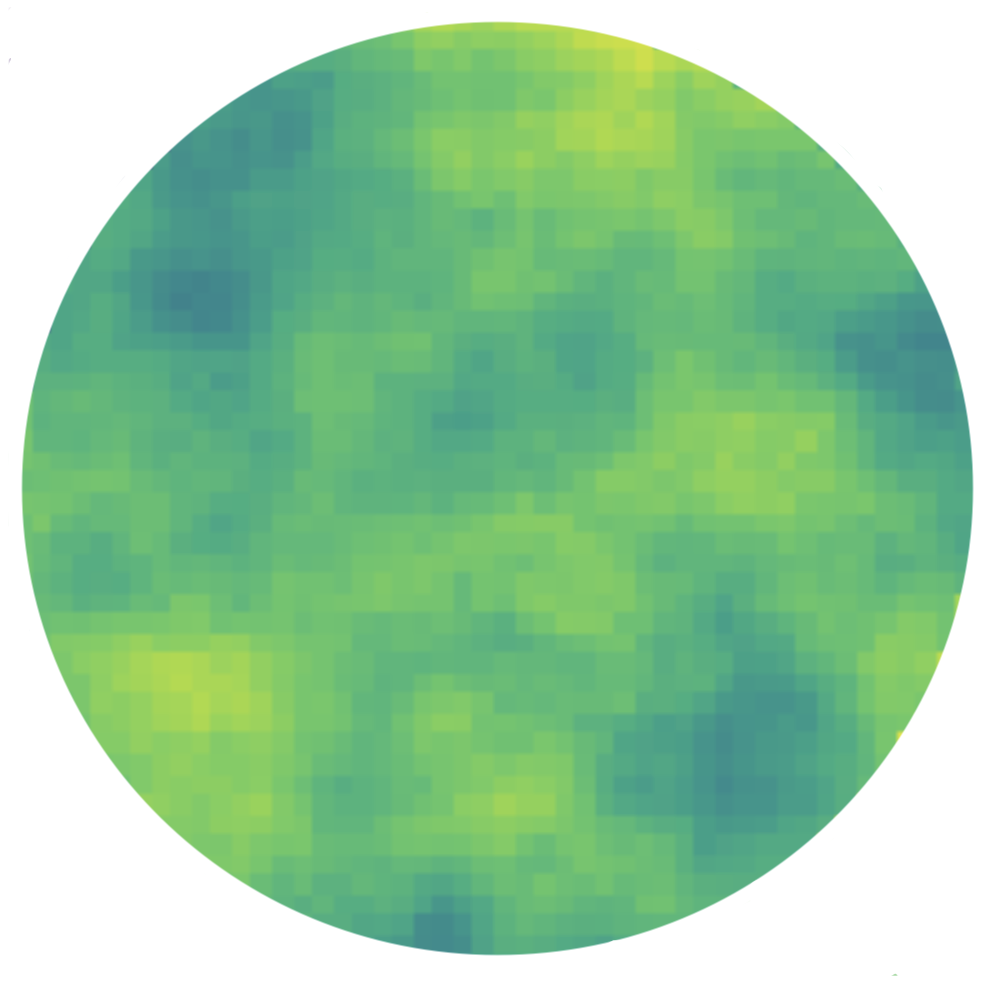}
   \hspace{2em}
   \includegraphics[height=0.32\linewidth]{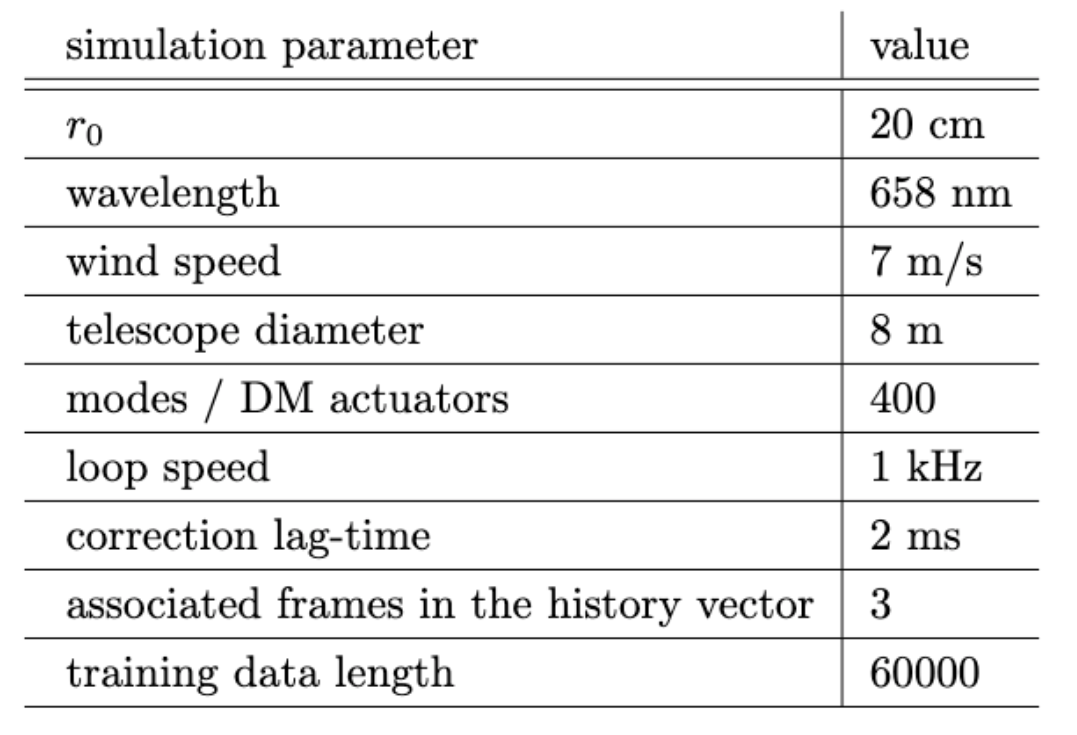}
   \end{tabular}
   \end{center}
   \caption[example] 
   { \label{fig:training_data} 
Left: Piston, tip, and tilt subtracted simulated phase screens from an 8 meter telescope with a single wind layer. Right: Table of parameters for generating the turbulence and AO simulations. }
   \end{figure}

{\label{sec:truth_conditions}}
The truth condition \textbf{P}, against which the predictive filter trains, must be the full uncorrected state of the wavefront (i.e., open loop turbulence) at a given iteration. The original work from Dessenne \cite{Dessenne1998} estimated the open loop wavefront from the transfer function of the AO control loop, essentially a higher fidelity pseudo open loop reconstruction than that of Guyon, 2017 \cite{Guyon2017}. For these simulations, we give the predictor perfect knowledge, using the full scale of turbulence from simulation. We note that this training condition is not realistic to on-sky operation, but acts as a laboratory to test the perfect performance of a closed-loop implementation, and future work will explore more realistic \textbf{P} generation. Figure \ref{fig:training_data} shows the example data used as the truth condition for training the closed-loop implementation. 

\section{PRELIMINARY IMPLEMENTATION OF CLOSED-LOOP PREDICTIVE CONTROL}
\label{sec:results}
We simulate atmospheric phase screens and an idealized AO control system with \texttt{HCIPy}\cite{hcipy}. We estimate the root-mean-square (RMS) residual error across the pupil for 1) the full state of the uncorrected turbulence, 2) a pseduo-integrator, in which we apply a perfect correction with a 2 ms time delay (for Keck, the time-delay is $\sim$1.5 ms\cite{Cetre2018}), 3) open loop EOF (well optimized in previous work \cite{jensen2019, Fowler2022}), and 4) a preliminary implementation of closed-loop EOF. These simulations use perfect wavefront sensing and correction, and are meant to provide comparative estimates only for the bandwidth error. The parameters for the simulation are shown in Figure \ref{fig:training_data}.

   \begin{figure} [h!]
   \begin{center}
   \begin{tabular}{c} 
   \includegraphics[height=0.38\linewidth]{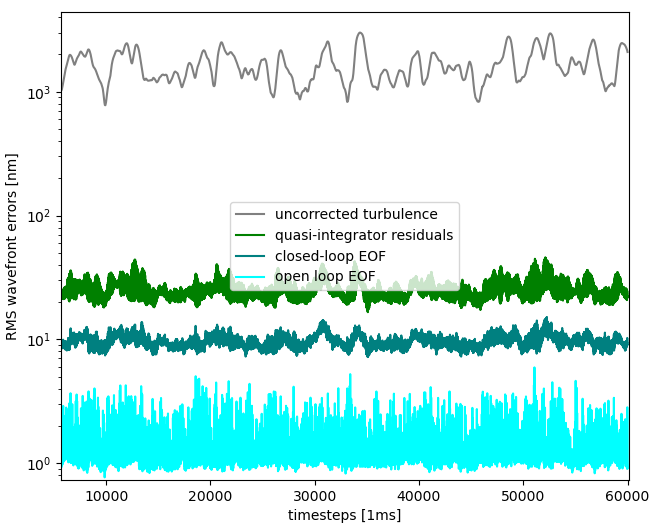}
    \includegraphics[height=0.38\linewidth]{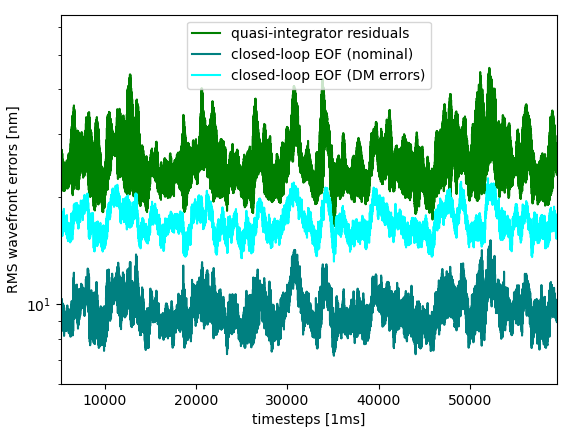}
   \end{tabular}
   \end{center}
   \caption[example] 
   { \label{fig:cl_vs_ol_rms} 
 Left: Two predictive methods alongside a psuedo-integrator (a perfect correction 2 timesteps behind), all applied with no measurement or fitting error (i.e., perfect wavefront sensing and control) running at 1 kHz. Full uncorrected turbulence RMS median: 1534.21 nm, quasi-integrator: 24.30 nm, closed-loop predictor: 9.49 nm, open loop predictor: 1.03 nm. Right: Impacts of a DM-model mismatch. Closed-loop EOF correction goes from a median of 9.49 (for the nominal run) to 16.66 nm RMS when a DM model error is introduced, but still shows better performance than the pseudo-integrator. The open loop EOF error increases by an order of magnitude and is not displayed on the figure for scale purposes.}
   \end{figure} 

\subsection{Performance in an Idealized AO System}

Figure \ref{fig:cl_vs_ol_rms} (left) shows the results of our initial implementation; we find a factor of improvement in RMS error of $\sim$ 2.5 over a standard integrator from closed-loop EOF, under-performing compared to the standard psuedo open loop. In practice, we should be able to recreate similar (if not better) performance between a closed and open loop implementation, however closed-loop minimization problem contains twice as many variables (treating the WFS and DM independently), twice as many regressors, and will require more training data to converge to an optimal solution, as shown in van Kooten, 2019 \cite{vanKooten2019}.  Future work will explore optimizing filter length and training length for a closed-loop implementation for a more fair comparison with open loop.

\subsection{Robustness to Model Errors}
Figure \ref{fig:cl_vs_ol_rms} (right) shows the performance of both predictors on a second set of data, in which we simulate a DM model mismatch in our control system -- adding a static factor of 2 between what the DM expects to apply and actually applies. We see that the performance of closed-loop EOF is less sensitive to errors in the system model. While the closed-loop performance drops when a DM model issue is introduced, the closed-loop predictor still outperforms a typical integrator. However, introducing the same issue into a open loop EOF increases the residual error by more than an order of magnitude. We note that in theory a closed-loop EOF should exactly learn and reconstruct the model-error, and future work to optimize the filter and training length will likely make a closed-loop EOF even more robust to model errors. 


\section{CONCLUSION}

In conclusion, we revisit the work of Dessenne, 1998 \cite{Dessenne1998} to compare open and closed-loop implementations of empirical orthogonal function (EOF) in simulation. A preliminary simulation of closed-loop EOF does not perform as well as an optimized implementation of open loop EOF, but still provides improvement over a classic integrator. A closed-loop implementation also proves to be more robust to the introduction of DM model errors.

We speculate that future optimization of the closed-loop predictive filter (e.g., exploring history vector length, training data length, etc) will likely close the gap between the open and closed-loop implementations and provide a controller that is even more robust to model errors. We also note that for our preliminary simulations we used perfect knowledge of the system to train our closed-loop predictive filter, which is not realistic for on-sky implementation; if performance comparisons prove promising, a more realistic training method could be devised, for example using the transfer function as in Desenne's original work \cite{Dessenne1998} or training on open and closed-loop data at the beginning of the night. Future work could also examine the comparative stability of open and closed-loop methods, as well as the impact of a more robust consideration of time-delay. 

We revisit this closed-loop predictive controller not only as an exploratory method for AO control, but also as a laboratory to explore model-mismatch and improve performance of the open loop implementations that are operating on-sky. For extreme adaptive optics on extremely large telescopes, novel ways to account for bandwidth error are worth pursuing. 

\appendix    

\section{TIME REFERENCE FRAMES AND TURBULENCE RECONSTRUCTION}\label{sec:time-steps}

\label{sec:time-steps}

One intriguing issue when comparing and reworking control methods is a robust understanding of time delays over the course of a control loop and how that plays out with different methods. Pseudo-open loop implementations gloss over this issue, by recreating a single point of information at a single point in time, but closed-loop implementations provide the opportunity to represent each piece of information at the time it enters the system, some work even accounts for fractional time-delays \cite{Poyneer2023}. The following is a minimal proof of information flow in a control loop, switching between time reference frames. 

   \begin{figure} [ht]
   \begin{center}
   \begin{tabular}{c} 
   \includegraphics[width=\linewidth]{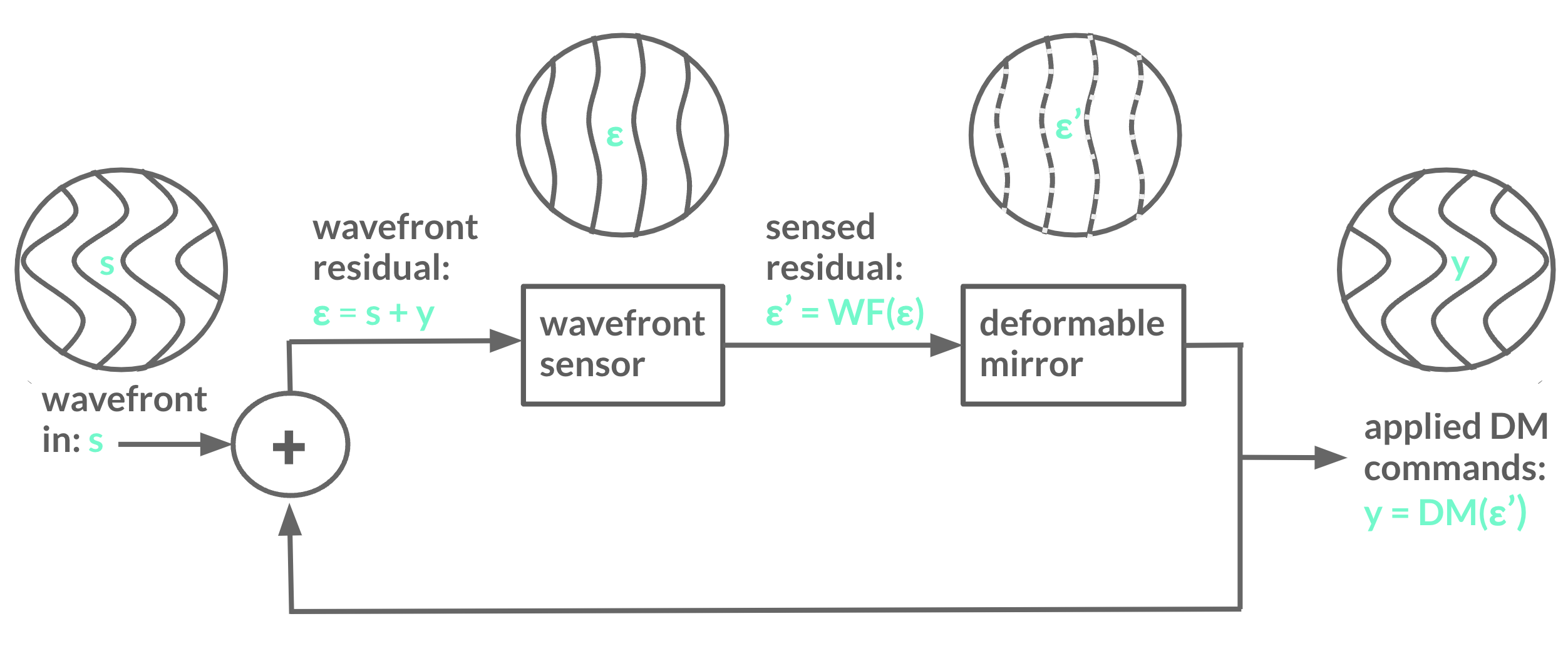}
   \end{tabular}
   \end{center}
   \caption[example] 
   { \label{fig:example} 
Control diagram of a classic two-step delay. The wavefront ($s$) comes into the system, and first meets a sum junction, where we apply deformable mirror (DM) commands ($y$), making it a closed-loop system. At this point, our residual is $\epsilon = s + y$. That signal is sensed by a wavefront sensor; that sensed signal is $\epsilon' = \textrm{WF}(\epsilon)$, where WF represents a functional form of how the wavefront sensor interprets the signal. Finally, this is fed into a computer that will calculate and apply a correction $y = DM(\epsilon')$, where DM represents a functional form of how that correction is applied and calculated based on $\epsilon'$.}
   \end{figure} 

We consider this system in two reference frames: (1) the time represented by the physical information flowing through the system (i.e., what time the turbulence the system is analyzing occurred); this is physically intuitive for timekeeping and residual comparison and we call it a physical clock and (2) the time represented by the control system and when information could be sampled from various sensors or correctors, which we call a control clock. 

Dessenne 1998 \cite{Dessenne1998}, uses a control clock framework. If, for example, we wanted to associated 3 states in time, we would build a history vector (used to predict the state at iteration n) of the form: 

\begin{equation}
\label{eq:phi}
\boldsymbol{\phi}(n) = 
\begin{bmatrix}
y(n-1) \\
y(n-2) \\
y(n-3) \\
\epsilon'(n-2) \\
\epsilon'(n-3) \\
\epsilon'(n-4) 
\end{bmatrix}
\end{equation}

The goal of this appendix is to show that these time steps from the perspective of the control clock associate logical pieces of information from the perspective of a physical clock. If we consider information moving through a system with a two step delay, we would see the chain of control events outlined in Table \ref{tab:time-clock}

\begin{table}[ht]
\caption{How information moves through the control loop, starting from the loop turning on. With a step of delay between each component, we at first have no information sensed and turned into a DM command. } 
\label{tab:time-clock}
\begin{center}       
\begin{tabular}{l|lll} 
\hline
\rule[-1ex]{0pt}{3.5ex}  $s$ & $\epsilon$ & $\epsilon'$ & $y$  \\
\hline
\rule[-1ex]{0pt}{3.5ex}  $s_1$ & $\epsilon_1 = s_1$ & - & - \\
\hline
\rule[-1ex]{0pt}{3.5ex}  $s_2$ & $\epsilon_2 = s_2$ & $\epsilon'_2 = \textrm{WF}(\epsilon_1)$ & - \\
\hline 
\rule[-1ex]{0pt}{3.5ex}  $s_3$ & $\epsilon_3 = s_3 + y_3$ & $\epsilon'_3 = \textrm{WF}(\epsilon_2)$ & $y_3 = \textrm{DM}(\epsilon'_2)$\\
\hline 
\rule[-1ex]{0pt}{3.5ex}  $s_4$ & $\epsilon_4 = s_4 + y_4$ & $\epsilon'_4 = \textrm{WF}(\epsilon_3)$ & $y_4 = \textrm{DM}(\epsilon'_3)$ \\
\hline 
\rule[-1ex]{0pt}{3.5ex}  $s_5$ & $\epsilon_5 = s_5 + y_5$ &$\epsilon'_5 = \textrm{WF}(\epsilon_4)$ & $y_5 = \textrm{DM}(\epsilon'_4)$ \\
\hline
\rule[-1ex]{0pt}{3.5ex}  $s_6$ & $\epsilon_6 = s_6 + y_6$ & $\epsilon'_6 = \textrm{WF}(\epsilon_5)$ & $y_6 = \textrm{DM}(\epsilon'_5)$ \\
\hline
\rule[-1ex]{0pt}{3.5ex}  $s_7$ & $\epsilon_7 = s_7 + y_7$ & $\epsilon'_7 = \textrm{WF}(\epsilon_6)$ & $y_7 = \textrm{DM}(\epsilon'_6)$ \\
\hline
\rule[-1ex]{0pt}{3.5ex}  $s_8$ & $\epsilon_8 = s_8 + y_8$ & $\epsilon'_8 = \textrm{WF}(\epsilon_7)$ & $y_8 = \textrm{DM}(\epsilon'_7)$ \\
\hline
\end{tabular}
\end{center}
\end{table}

If we now consider the history vector $\boldsymbol{\phi}$, we can rewrite it to prediction some iteration $n=7$

\begin{equation}
\boldsymbol{\phi}(7) = 
\begin{bmatrix}
y(n-1) = y_{6} = \textrm{DM}(\epsilon_5') = \textrm{DM(WF}(\epsilon_4)) \\
y(n-2) = y_{5} = \textrm{DM}(\epsilon_4') = \textrm{DM(WF}(\epsilon_3))\\
y(n-3) = y_{4} = \textrm{DM}(\epsilon_3') = \textrm{DM(WF}(\epsilon_2)) \\
\epsilon'(n-2) = \epsilon_5' = \textrm{WF}(\epsilon_4) \\
\epsilon'(n-3) = \epsilon_4' = \textrm{WF}(\epsilon_3) \\
\epsilon'(n-4) = \epsilon_3' = \textrm{WF}(\epsilon_2)
\end{bmatrix}
\end{equation}

Notice that though the control clock indexing appears to introduce a one step offset, the most expanded part of the expression for each iteration shows that the elements on the history vector depend on the the same physically timed pieces of information. 


It should be noted that a classic two-step delay is not actually physically reminiscent of most systems. In Cetre, 2018\cite{Cetre2018} they found that for the Keck pyramid WFS RTC, the time delay over the entire correction (including wavefront sensing time, calculations, and DM latency after correction is applied) takes $\sim$1.5 ms, where $\sim$1 ms is the wavefront sensing and readout time, and $\sim$0.5 ms is the calculation and hardware latency time. For this simulation work we have opted to only update the DM once per new piece of wavefront sensor information, essentially forcing this to be a classic two-step delay. However, future work will explore the efficacy of updating the DM more frequently, as we could project forward the correction easily and provide two DM updates per wavefront sensor readout.

\printbibliography 
\end{document}